\documentclass[useAMS,usenatbib,usegraphicx]{mn2e}

\title[The planetary system host HR\,8799: On its $\lambda$ Bootis nature]{The planetary system host HR\,8799: On its $\lambda$ Bootis nature}
\author[A. Moya et al.]{A. Moya$^{1}$\thanks{E-mail:
amoya@cab.inta-csic.es}, P. J. Amado$^{2}$, D. Barrado$^{1}$,
  A. Garc\'{\i}a Hern\'andez$^{2}$, \and M. Aberasturi$^{1}$,
  B. Montesinos$^{1}$, F. Aceituno$^{2}$\\
$^{1}$Departamento de Astrof\'{\i}sica, Laboratorio de Astrof\'isica
  Estelar y Exoplanetas, LAEX-CAB (INTA-CSIC), PO BOX 78,\\
 28691 Villanueva de la Ca\~nada, Madrid, Spain\\
$^{2}$Instituto de
  Astrof\'{\i}sica de Andaluc\'{\i}a - CSIC, Camino Bajo de Huetor 50,
  18008, Granada, Spain}
\begin{document}



\maketitle

\label{firstpage}

\begin{abstract}
HR\,8799 is a $\lambda$ Bootis, $\gamma$ Doradus star hosting a
planetary system and a debris disk with two rings. This makes this
system a very interesting target for asteroseismic studies. This work
is devoted to the determination of the internal metallicity of this
star, linked with its $\lambda$ Bootis nature (i.e., solar
  surface abundances of light elements, and subsolar surface
  abundances of heavy elements), taking advantage of its $\gamma$
Doradus pulsations. This is the most accurate way to obtain this
information, and this is the first time such a study is performed for
a planetary-system-host star. We have used the equilibrium code CESAM
and the non-adiabatic pulsational code GraCo. We have applied the
Frequency Ratio Method (FRM) and the Time Dependent Convection theory
(TDC) to estimate the mode identification, the Brunt-Va\"is\"al\"a
frequency integral and the mode instability, making the selection of
the possible models. When the non-seismological constraints (i.e its
position in the HR diagram) are used, the solar abundance models are
discarded. This result contradicts one of the main hypothesis for
explaining the $\lambda$ Bootis nature, namely the accretion/diffusion
of gas by a star with solar abundance. Therefore, according to these
results, a revision of this hypothesis is needed. The inclusion of
accurate internal chemical mixing processes seems to be necessary to
explain the peculiar abundances observed in the surface of stars with
internal subsolar metallicities. The use of the asteroseismological
constraints, like those provided by the FRM or the instability
analysis, provides a very accurate determination of the physical
characteristics of HR\,8799. However, a dependence of the results on
the inclination angle $i$ still remains. The determination of this
angle, more accurate multicolour photometric observations, and high
resolution spectroscopy can definitively fix the mass and metallicity
of this star.
\end{abstract}


\section{Introduction}

The A5 V star HR\,8799 (V342 Peg, HD\,218396, HIP\,114189) has been
extensively studied in the last years. The first studies of this star
were done in the context of asteroseismology. This technique is being
developed as an efficient instrument for the study of stellar
interiors and evolution \citep{libroastro}. Through the comparison of
the observed pulsational modes with theoretical calculations, the
general and internal characteristics of the star can be
determined. \cite{schuster} firstly reported HR\,8799 as a possible SX
Phoenicis type. \cite{zerbi} observed this star in a multisite
multicolour photometric campaign, with Str\"omgren filters, and found
three independent pulsational frequencies
($f_1=1.9791\;\rm{c}\,\rm{d}^{-1}$ ($\equiv$ 22.906 $\mu$Hz),
$f_2=1.7268\;\rm{c}\,\rm{d}^{-1}$ ($\equiv$ 19.986 $\mu$Hz), and
$f_3=1.6498\;\rm{c}\,\rm{d}^{-1}$ ($\equiv$ 19.095 $\mu$Hz), units are
cycles per day), making it one of the 12 first $\gamma$ Doradus
pulsators known \citep{kaye1}. This pulsating stellar group is
composed of Main Sequence (MS) stars in the lower part of the
classical instability strip \citep{tdcma}. Their pulsation modes have
periods in the range [0.5,3] days, that is, they are asymptotic g-mode
pulsators. \cite{gray} obtained an optical spectrum of HR\,8799, and
assigned an spectral type of kA5 hF0 mA5 V $\lambda$ Bootis, reporting
an atmospheric metallicity of [M/H]=$-0.47$. They also noted
that HR\,8799 may be also a Vega-type star, characterized by a far IR
excess due to a debris disk. \cite{sadakane} developed a deep study of
the metal abundances of this star, confirming its $\lambda$ Bootis
nature (with surface chemical peculiarities).

Two years later, \cite{Marois} reported the presence of a planetary
system around this star. It was the first detection of such a system
carried out by direct imaging. This was the starting point of a set of
approximately ten studies about this system during 2009, none of them
from the asteroseismic point of view, even considering that the three
detected pulsational frequencies can be used to better understand the
host star, in particular its $\lambda$ Bootis nature and evolutionary
status.

Discovered by \citet{Morgan43}, the $\lambda$ Bootis-type stars are
non-magnetic, moderately-rotating, Population I stars with spectral
types from late B - early A to F (dwarfs)\footnote{Two percent of the
  A-type stars belong to this class \citep{GrayCorbally02}}, which
show peculiarities in the morphology and abundance of the Fe-peak
element lines. In particular, these lines are unusually weak
considering their spectral types. Significant deficiencies in their
abundances (up to 2\,dex) are found, whereas C, N, O and S have solar
abundance \citep{Paunzeniliev02}. Different theories have attempted to
explain the $\lambda$ Bootis nature from both observational
(photometry, spectroscopy) and theoretical investigations. It is not
our aim to discuss all of them here \citep[for a interesting review
  see][]{Paunzen03}. Nevertheless, it may be worth describing the most
accepted scenario relying on the accretion of inter-stellar medium gas
by the star \citep{VennLambert90}. The accretion/diffusion scenario
would explain the abundances found at the base of the outer convective
zone of these stars, since convective layers are assumed to remain
chemically homogeneous. This accretion/diffusion scenario is
  based on the accretion of inter-stellar medium by the star, and the
  mixing of these elements with those of the star due to diffusion and
  rotationally mixing processes. The accretion rate required to
maintain this situation is of the order of $10^{-10}$-$10^{-14}\,{\rm
  M}_{\odot}$ per year \citep{turcotte}, and, once the accretion has
ceased, the metal deficiencies should disappear in approximately 1 Myr
due to diffusion and internal mixing processes. Thus, a possible
interpretation is that $\lambda$ Bootis stars are young A-type stars
(in a pre-main sequence or zero-age main sequence evolutionary stage),
still interacting with their primordial clouds of gas and dust.
Interestingly, \citet{Paunzeniliev02} found that most of the known
$\lambda$ Bootis stars lie between the ZAMS (zero-age main sequence)
and the TAMS (terminal-age main sequence, with ages of several
hundreds million years). In this case the most likely scenario would
be a MS star with solar abundance passing through an interstellar
cloud \citep{kamppaunzen}. However, the chemical mixing due to
internal processes, such as rotationally-induced mixing, cannot be
discarded as a possible explanation of the observed abundances.

$\lambda$ Bootis stars and other types of objects in the same region
of the HR diagram, such as $\delta$ Scuti and $\gamma$ Doradus
stars, are considered as particularly suitable for the
asteroseismological study of poorly known hydrodynamical processes
occurring in stellar interiors, like the extent of the convective
core, mixing of chemical elements, redistribution of angular momentum
\citep{Zahn92,steph1,steph2}, etc. $\lambda$ Bootis-type stars are
also pulsating stars. Therefore, asteroseismology can be used to
obtain information about the internal structure of these
objects. Several works have been devoted to study $\lambda$ Bootis
stars with $\delta$ Scuti pulsations, for instance
\cite{Paunzen98,Casas09}. In addition, the combined use of space and
ground-based observations improves the potential of the modelling of
the star \citep{bruntt}.

The present work aims at a comprehensive asteroseismic modelling of
HR\,8799, focusing on the discussion of the $\lambda$ Bootis nature of
the star. More precisely, we want to answer whether the observed
abundance is intrinsic to the star or an effect of surface
processes. This answer drives the search for possible mechanisms
explaining the observed abundances, that is, if there is accretion or
not, together with internal chemical transport (rotationally-induced
mixing, gravitational settling, radiative levitation, etc.). Up to
now, only three $\lambda$ Bootis stars have been reported to be
$\gamma$ Doradus pulsators: HD\,218427 \citep{rodriguez06a},
HD\,239276 \citep{rodriguez06b}, and HR\,8799. As $\gamma$ Doradus
stars are asymptotic g-mode pulsators, they are very good candidates
for testing the internal structure of the star, in particular its
internal metallicity. Some of the most updated tools adapted for this
purpose are used: 1) the evolutionary code CESAM \citep{Morel08}, and
2) the pulsation code GraCo \citep{graco1,graco2}. Using these tools
we performed a massive numerical study of HR\,8799 in an attempt of
constraining physical and theoretical parameters. In this work we will
follow the same scheme used for the study of RV Arietis, 29 Cygnis and
9 Aurigae \citep{Casas06,Casas09,Moya06au}.









\section{Observations and the properties of HR\,8799}

The $\lambda$ Bootis star HR\,8799 was discovered to be a pulsating
$\gamma$ Doradus star by \cite{zerbi}. They observed it using
multicolour Str\"omgren photometry and found three frequencies with
their corresponding amplitudes and phases in the different wavelenght
bands. In Table 1 these frequencies, and the amplitudes and
  phases for the two highest amplitude frequencies, are
  shown. \cite{gray} obtained spectroscopic observations of this
star, providing accurate values for the stellar luminosity,
$T_{\rm{eff}}$, and $\log\,g$ (see Table 2). They also
established the $\lambda$ Bootis and Vega-type nature of this star. In
a study by \cite{sadakane}, a more detailed spectroscopic analysis of
this star was done. An accurate determination of the individual
atmospheric abundances was obtained, confirming the previous
results. In a recent paper, \cite{cuypers} carried out a long-term
photometric monitoring of a set of $\gamma$ Doradus stars, including
HR\,8799. The effective temperature reported in that work is similar
to that given in \cite{gray}.

An interesting additional property of HR\,8799 was the observation, by
HST and Spitzer, of two well separated rings in its debris disk,
located at the inner and the outer limits of the planetary system
\citep{chen,reider}.

\subsection{New spectroscopic observations}

We have obtained intermediate resolution, high signal-to-noise
spectra in the range 3600--7800 \AA, with CAFOS (Calar Alto Faint
Object Spectrograph) on the 2.2-m telescope at Calar Alto Observatory
(CAHA, Almer\'{\i}a, Spain). The observations were taken on 9 November
2009. CAFOS was equipped with a CCD SITe detector of 2048$\times$2048
pixels (pixel size 24 $\mu$m) and the grisms Blue-100 and Green-100,
giving a linear dispersion of $\sim 88$ \AA/mm (2 \AA/pixel).

Spectroscopic observations in the range 3600--6800 \AA{} were also
obtained on 12 November 2009 with the spectrograph Albireo on the
1.5-m telescope at the Sierra Nevada Observatory (OSN, Granada,
Spain). This instrument is equipped with a CCD2k detector of
2048$\times$2048 pixels. The grating 1200 was used, providing a linear
dispersion of $\sim 67$ \AA/mm (1 \AA/pixel).

In both cases the slit width was 1.5 arcsec. The usual bias, dark,
dome flat-field and calibration lamp frames were taken. Standard
procedures were used to process the data. The widths of the lines of
the calibration lamps were used to obtain the instrumental
responses of the spectrographs.

\subsection{Spectral fitting and properties}

The aim of the observations was to carry out a double check of
some of the stellar parameters collected from the literature. In
Fig. 1 we show the CAFOS spectrum of HR 8799 in the interval
4000--5000 \AA\, (containing the H$\beta$, H$\gamma$ and H$\delta$
Balmer lines) plotted as a black solid line, superimposed by a
synthetic spectrum computed using $T_{\rm eff}\!=\!7430$ K, $\log
g\!=\!4.35$, [M/H]=$-0.5$ and $v \sin i$=37.5 km s$^{-1}$. The
resolution of the synthetic spectrum has been degraded to match that
of the observed one. The agreement between both spectra is quite
remarkable, indicating that the stellar parameters used are
reliable.

The spectral synthesis has been done using the ATLAS9 and SYNTHE codes
by \cite{kur93}.  In this work we have used the GNU Linux version of
the codes available on-line\footnote{{\tt
    http://wwwuser.oat.ts.astro.it/atmos/}} \citep{sbordone}. Model
atmospheres provided by \cite{kurucz} are used as one of
the inputs for the synthesis.

\begin{figure}
\includegraphics[width=84mm]{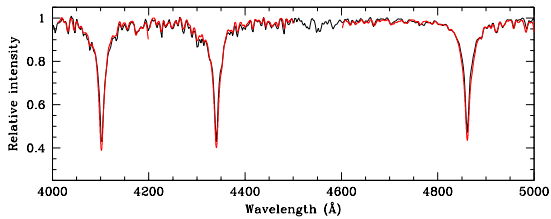}
 \caption{Spectrum of HR 8799 in the interval 4000--5000 \AA{} (black)
   and, superimposed, a synthetic spectrum computed with $T_{\rm
     eff}\!=\!7430$ K, $\log g\!=\!4.35$, [M/H]=$-0.5$ and $v \sin
   i$=37.5 km s$^{-1}$ (red), based on model atmospheres by
     Kurucz.}
\end{figure}

Fig. 2 shows the position of HR\,8799 in the colour magnitude diagram
(CMD). Evolutionary tracks are from \cite{claret}.  The main sequence
has been taken from \cite{philip} and the $\delta$ Scuti instability
strip from \cite{robre}.  The $\gamma$~Dor instability strip is from
\cite{handsho}. Solid stars represent bona fide $\gamma$ Doradus
objects observed by \cite{handsho}, and solid triangles are bona fide
$\gamma$ Doradus stars from the literature. The solid and open
circles are prime and less probable $\gamma$ Doradus candidates
respectively from those authors. We have derived photometrically
$M_{\rm{V}}$ and $(b-y)_0$ using Str\"omgren data obtained from the
Hauck-Mermilliod catalogue \citep[;solid star]{Hauc98}. The reddening
correction was negligible. On the other hand, using the Hipparcos
parallax ($25.38\pm0.85$ mas, i.e., a distance of 39.40 pc), an
absolute magnitude $M_V$=5.96 was calculated. The Hipparcos
$M_{\rm{V}}$ is plotted as an open star. Other available parallaxes in
the literature do not change significantly the absolute magnitude of
this star (less than 0.1 $L/L_{\odot}$).

The use of different scales of bolometric corrections yields
significant larger luminosities (up to 0.3 $L_\odot$), compromising
the study presented here. We have used the most updated tool provided
by Virtual Observatory, VOSA \citep{amelia} to fit Kurucz atmospheric
models to all the photometric observational data available in the
literature, avoiding the necessity for estimating the stellar
luminosity using bolometric corrections. Models with realistic
metallicities, best fitting observations, provide a luminosity similar
to that given in \cite{gray} (about 0.1 $L/L_{\odot}$ larger).

\begin{figure}
\includegraphics[width=84mm]{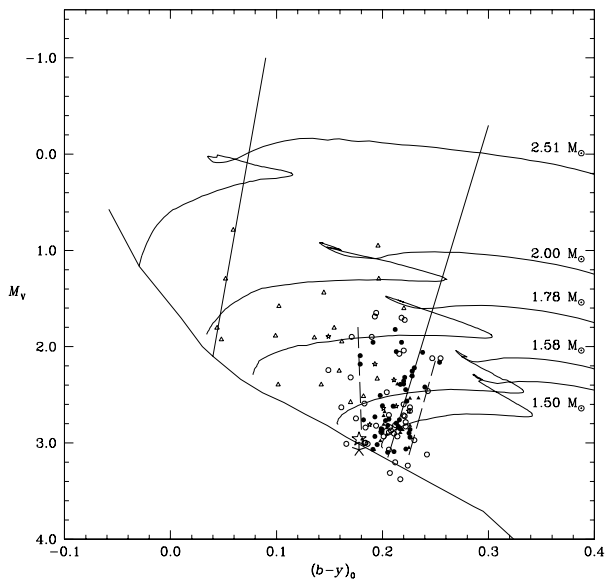}
 \caption{Colour magnitude diagram with absolute visual magnitudes
   $M_{\rm{V}}$ plotted against dereddened Str\"omgren ($b-y$)$_0$
   colour. Small symbols are from \protect\cite{handsho}: solid stars
   represent bona fide $\gamma$ Doradus stars observed in their work,
   and solid triangles are bona fide $\gamma$ Doradus stars from
   the literature. The small solid and open circles are prime and less
   probable $\gamma$ Doradus candidates respectively. The position of
   HR\,8799 is determined by using its Str\"omgren magnitudes taken
   from the Hauck-Mermilliod catalogue \protect\citep{Hauc98} and its
   absolute visual magnitude. Two values were derived for the latter,
   one derived from the Str\"omgren photometry using the calibrations
   by \protect\cite{Craw75} and \protect\cite{Craw79}, implemented in
   the program TempLogG \protect\citep{Roge95} and another one using
   the Hipparcos parallax.  These two determinations are given in the
   plot by the larger solid and open star symbols respectively.}
\end{figure}

\begin{table*}
 \centering
 \begin{minipage}{140mm}
  \caption{Str\"omgren multicolour photometric
      observations. Phase differences and amplitude ratios of filters
      $u$, $v$ and $b$ related to filter $y$ for the two
      frequencies with highest amplitude are also shown.}
  \begin{tabular}{rrrrrrr}
  \hline
 freq  &  $u-y$ & $v-y$ & $b-y$ &  $u/y$ &   $v/y$  &  $b/y$\\
 (c/d) &  $(^\circ)$ & $(^\circ)$ & $(^\circ)$ &&&\\
 \hline
1.9791 ($\equiv$ 22.906 $\mu$Hz) & -0.5$\pm$3.6 & 2.5$\pm$2.8 & 1.8$\pm$2.8 & 0.936$\pm$0.057 &
1.470$\pm$0.071 & 1.275$\pm$0.061\\
1.7268 ($\equiv$ 19.986 $\mu$Hz) & -7.0$\pm$7.5 & -0.8$\pm$5.7 & -1.4$\pm$5.7 & 0.830$\pm$0.112
& 1.316$\pm$0.137 & 1.181$\pm$0.131\\
1.6498 ($\equiv$ 19.095 $\mu$Hz) & - & - & - & - & - & -\\
\hline
\end{tabular}
\end{minipage}
\end{table*}

\begin{table*}
 \centering
 \begin{minipage}{130mm}
  \caption{Physical characteristics of HR\,8799 taken from: 
    1) \protect\cite{gray}
    2) \protect\cite{kaye2}
    3) \protect\cite{leeuwen}}
  \begin{tabular}{rrr}
 \hline
$T_{\rm{eff}}$ (K) & 7430$\pm$75 & ref. 1\\
$\log\,g$ ($\rm{cm}\,\rm{s}^{-2}$) & 4.35$\pm$0.05 & ref. 1\\
$M_{\rm{V}}$ & 2.98$\pm$0.08 & ref. 1\\
$R (R_\odot)$ & 1.34$\pm$0.05 & ref. 1\\
$L (L_\odot)$ & 4.92$\pm$0.41 & ref. 1\\
$v\sin\,i$ (km $\rm{s}^{-1}$) & 37.5$\pm$2 & ref. 2\\
$\pi$ (mas) & 25.38$\pm$0.85 & ref. 3\\
\hline
\end{tabular}
\end{minipage}
\end{table*}

\begin{figure}
\includegraphics[width=84mm]{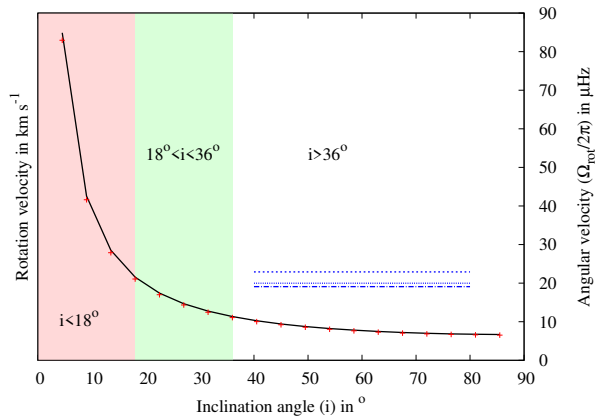}
 \caption{Rotation and angular velocities of HR\,8799 as a function of
   the inclination angle $i$. The red region ($i<18^\circ$) represents
   the zone where the star can hardly be a $\lambda$ Bootis star. The
   green region ($18^\circ <i<36^\circ$) represents the zone where our
   procedure cannot be applied accurately. The horizontal lines are
   the observed frequencies of HR\,8799.}
\end{figure}


\subsection{Rotation and the limits of our analysis}

An important observational uncertainty for the present study comes
from the inclination angle $i$ between the rotation axis of the star
and the observer line of sight. Using the rotation velocity $v\sin
i=37.5\,\rm{km}\,\rm{s}^{-1}$ given by \cite{kaye2}, several works in
the literature tried to provide estimations for $i$, as in
\cite{reider}, where one of the detected frequencies have been used as
signature of the rotation velocity, obtaining three estimations for
the inclination angle, one per observed frequency. The amplitude and
phase differences observed in different colour bands for these
frequencies make this interpretation invalid. Other
  determinations of the inclination angle in the literature are
  related with the observation of the debris disk and/or the dynamics
  of the planetary system. \cite{fabrycky} found a lower limit
  $i>20^\circ$ based in the dynamical stability of the system, assuming
  the mass of the planets provided by \cite{Marois}. On the other
  hand, \cite{su} studied the debris disk of HR\,8799 and found that
  inclination angles larger than $25^\circ$ must be ruled out. We have to
  take this latter result with caution since we must take into account
  that the morphology inferred from the observations at 70
  $\mu$m is not a good constraint for the inclination angle of the
  star. That emission is dominated by a distribution of small grains
  and the geometry and the inclination angle we derive would
  correspond to those of disk-like distribution of such
  particles. That inclination angle may not coincide with the
  inclination angle of the star \citep{bate}.

In the present study, we use the Frequency Ratio Method (FRM)
developed in \cite{frm} (see Section 3.2) for the asteroseismological
modelling of this star. This procedure has a limited range of validity
in stellar rotation velocity, as demonstrated in \cite{frmrot}. The
rotation velocities used in this work and the pulsational frequencies
are around 0.1 and 0.25 of the break-up rotation velocity
respectively. Therefore, the use of a second order perturbation theory
for the calculation of the frequencies with rotation is accurate
enough \citep{ballot}. That second order theory is used for the study
done by \cite{frmrot}. Nevertheless, as the observed frequencies are
larger but close to twice the angular rotation velocities used in this
work, the use of the traditional approximation to study the accuracy
of the asymptotic expression in the presence of rotation would be
recommended \citep{lee,townsend,mathis}. This study needs to be
carried out to determine more precisely the reliability of the FRM.

In \cite{frmrot}, the authors found a maximum rotation velocity of
around $60\,\rm{km}\,\rm{s}^{-1}$ for the correct application of the
FRM, for a standard $\gamma$ Doradus star. On the other hand,
\cite{turcotte} found that meridional circulation cannot destroy the
accretion pattern (in the accretion/diffusion scenario) for a rotation
velocity below $125\,\rm{km}\,\rm{s}^{-1}$. These values provide two
lower limits for the inclination angle, one limited by the $\lambda$
Bootis nature and another one to ensure the applicability of the FRM.

In Figure 3, the rotation velocity as a function of $i$ for a stellar
radius of $R=1.30\,R_{\odot}$ is shown. For the different stellar
radii determined for this star the results are similar. In this figure
we see in red the region with $i<18^\circ$, limit imposed by the
$\lambda$ Bootis nature of HR\,8799. In green it is the region with
$18^\circ <i<36^\circ$, the limit for the FRM to be applicable. We
want to note that if the inclination angle lies in the green zone, our
analysis would be inaccurate. Finally, the observed pulsational
frequencies are displayed with horizontal lines. In this figure we see
how those frequencies are, at least, twice the rotation frequencies
used in this study.

\section{Tools and methods}

In this section, we give a brief description of the tools used for the
present study.

\subsection{Stellar equilibrium models}

The stellar equilibrium models were computed using the evolutionary
code CESAM, with a mesh grid (B-splines basis) of 2000
points. First-order effects of rotation on the equilibrium models were
considered by subtracting the spherically averaged contribution of the
centrifugal acceleration to the gravity of the model,
$g_{\rm{eff}}=g-{\cal A}_c(r)$, where $g$ corresponds to the local
gravity, and ${\cal A}_c(r)$ represents the radial component of the
centrifugal acceleration. This spherically averaged component of the
centrifugal acceleration does not change the order of the hydrostatic
equilibrium equations. Such models are called 'pseudo-rotating' models
\citep[see][]{Soufi98, Sua07neadeg}. During evolution, models are
assumed to rotate as a rigid body, and their total angular momentum is
conserved. Although the non-spherical components of the centrifugal
acceleration were not considered, they were included as a perturbation
in the computation of the oscillations.

Standard physical inputs for $\gamma$ Doradus stars are used, i.e.,
the CEFF equation of state \citep{ceff}. The opacity tables were taken
from the OPAL package \citep{Igle96}, complemented at low temperatures
($T\leq10^4\,\rm{K}$) by the tables provided by
\citet{AlexFergu94}. The atmosphere was calculated following two
approaches: a grey atmosphere (Eddington $T(\tau)$ law) when the
equilibrium models were used to compute adiabatic oscillations, and
the Kurucz model atmospheres for the computation of non-adiabatic
quantities. The abundance mixture used is that given in
\cite{Grevessenoels93}.

The main approximation taken in the models with a possible influence
in the results of this study is the lack of updated internal chemical
transport mechanisms in the equilibrium models, such as
rotation-induced mixing or gravitational waves. The inclusion of these
mechanisms is the next step to be taken, considering the conclusions
obtained in the present work.

\subsection{Oscillation computations}

The seismic models were completed by computing for each equilibrium model
its corresponding oscillation spectrum and non-adiabatic observables.

GraCo \citep{graco1,graco2} provides non-adiabatic quantities related
to pulsation and includes the convection - pulsation interaction using
the Time Dependent Convection theory \citep{tdcma}. This theory
improves the ``frozen convection'' approximation usually implemented
in most of the codes. With this we can study the instability of the
modes. It also provides accurate variations of the effective
temperature, effective gravity and phase-lags of the modes. These
quantities are the base of the mode identification using multicolour
photometry (see Section 3.5). The adiabatic solutions of this code
have been used as reference for the ESTA works \citep[Evolution and
  Asteroseismic Tools Activities,][]{estades,estafreq}.

\subsection{The Frequency Ratio Method}

The Frequency Ratio Method or FRM \citep{frm} takes advantage of
the analytical expression of the frequencies given by the asymptotic
theory \citep{tassoul,smeyers}. $\gamma$ Doradus stars and other slow
pulsators present frequencies in this g-mode asymptotic region. This
makes it possible to obtain information of the radial order $n$ of the
modes and the Brunt-V\"ais\"al\"a frequency integral of the model,
through the ratios of the observed frequencies. With at least three
frequencies the method can be carried out.

The application of the method to the three observed frequencies
provides chains of
$(n_1,\ell_1,n_2,\ell_2,n_3,\ell_3,\rm{I}_{\rm{obs}})$,
where $n_i$ is the radial order of mode $i$, $\ell_i$ is its
spherical order, and $\rm{I}_{\rm{obs}}$ is the new observable introduced by the
FRM, the observed Brunt-V\"ais\"al\"a frequency integral, the same for
the three frequencies since it is a global property of the stellar model.

\cite{frm} showed that the real solution is part of the set of
possible chains provided by the FRM. In addition, in this work we have
relaxed the requirement of all the modes having the same $\ell$. We
allow $\ell$ to take values 1 or 2, the most probable ones for
modes detected with ground-based photometry for $\gamma$ Doradus stars.

\cite{frmrot} showed the application limits of the FRM due to stellar
rotation. With the assumption of all the frequencies having the same
azimuthal order $m$, the main conclusions of that paper were: 1) the
rotation limit for a standard $\gamma$ Doradus star for the
perturbative approximation to be valid is around
$60\;\rm{km}\,\rm{s}^{-1}$, 2) if that assumption is fulfilled, the
real solution is still part of the list of possible solutions provided
by the FRM, otherwise the FRM does not give any solution, and 3) the
accuracy of the asymptotic expressions decreases with the rotation
velocity, increasing the number of possible solutions. Nevertheless,
these conclusions and application limits of the FRM must be revised
under the traditional approximation (see section 2.3).

\subsection{Instability analysis}

The energy balance of the modes during a complete period is one of the
most useful information we can use. Analysing the energy balance of a
pulsational mode with its surroundings in a pulsation cycle we can
infer if the mode is stable (losses energy) or over-stable (gains
energy). To do this study accurately, the main physical processes
having influence on this energy balance must be implemented in the
codes. \cite{guzik} and \cite{tdcma}, showed that the position of
  the base of the the convective envelope is the key point driving
  $\gamma$ Doradus modes, since they are originated by the blocking of
  the radiative flux at this point. This can explain the location of
  the blue edge of the $\gamma$ Doradus instability strip. For these
stars, the Time Dependent Convection \citep[TDC]{tdcma,tdcahmed} is
the only set of equations predicting unstable modes in the observed
frequency ranges. This development has provided the first theoretical
predictions of instability strips for these stars.

\subsection{Multicolour photometric analysis}

In the following, multicolour photometry is used to provide constraints
on the physical characteristics of the star through additional information on
the degree $\ell$ of the spherical harmonic associated to each
observed pulsational frequency.

The linear approximation to non-radial flux variations of a pulsating
star was first derived by \citet{Dziem77}, and later reformulated by
\citet{BalonaStobie79} and \citet{Watson88}. Then, \citet{Garrido90}
showed that the $v$ and $y$ Str\"omgren bands can be used for
discriminating the degree $\ell$. The comparison of the numerical
solutions with the observations is based on non-adiabatic calculations
\citep[more details in][]{graco1}. In particular, pulsation is
non-adiabatic in the outer convective zone of these stars, where the
thermal relaxation time is of the order of the pulsation
period. Accurate determination of the eigenfunctions in these layers,
therefore, requires the use of a non-adiabatic description which
includes the already mentioned TDC. This procedure makes it possible
to relate multicolour photometric observables with such
eigenfunctions, allowing therefore a direct constraining on some
unknown physical parameters through the direct comparison with
observations.

All these tools can be combined to provide an accurate determination
of the general physical properties of HR\,8799, as described in the
next section.

\section{Analysis of the star}

The analysis of this star has been done in different steps, imposing
the observational constraints sequentially. The sequence is: 1)
Fulfill the physical characteristics given by spectroscopy
\citep{gray}, 2) fulfill the Brunt-V\"ais\"al\"a frequency integral
constraints provided by the FRM, 3) predict the observed frequencies
to be over-stable, and 4) identify the spherical degree $\ell$ of the
mode for the frequency with the highest amplitude by using the
observed multicolour photometry.

\subsection{Position in the HR diagram}

\begin{table*}
 \begin{minipage}{140mm}
  \caption{Acceptable models depending on the physical
    constraints. Each row shows the results obtained when a new
    constraint is added with respect to the previous one. The bottom
    row shows the results obtained when all the constraints are
    imposed to the less favorable case from the FRM point of view (see
    text for details). The mass and metallicity determinations are
    linked, that is, the first mass range is linked with the first
    metallicity shown, etc.}
  \begin{tabular}{lrrrr}
  \hline
   Constraint     &   Mass &  Metallicity & $\alpha_{\rm{MLT}}$ & Overshooting \\
   & (in $\rm{M}_{\odot}$) & (in [M/H]) & & \\
 \hline
HR position & [1.25,1.27],[1.32,1.35],[1.40,1.48] & -0.52, -0.32, -0.12
& 0.5, 1.0, 1.5 & 0.1, 0.2, 0.3\\
Previous + FRM & [1.25,1.27],[1.32,1.34],[1.40,1.48] & -0.52, -0.32, -0.12
& 0.5, 1.0, 1.5 & 0.1, 0.2\\
Previous + Inst & [1.32,1.33],[1.44,1.45] & -0.32, -0.12 & 1.5 & 0.1, 0.2\\
Previous + colours & 1.32 & -0.32 & 1.5 & 0.1, 0.2\\
Total in less favorable case & [1.32,1.33],[1.44,1.45] &
-0.32, -0.12 & 1.5 & 0.1, 0.2, 0.3\\
\hline
\end{tabular}
\end{minipage}
\end{table*}

The first step of our study was to search for models in a dense grid
fulfilling the observed physical characteristics of the star. As it
has been explained in Section 2, we have used the effective
temperature, luminosity, and gravity (with their errors)
given in Table 2. The internal metallicity has been regarded as a
free parameter to test the $\lambda$ Bootis nature of the star and to
obtain the internal metallicity using asteroseismology. The mass,
estimated in $1.47\pm0.30\,\rm{M}_{\odot}$ by \cite{gray}, has been
also regarded as a free parameter since it has not been directly
determined.

\begin{figure}
\includegraphics[width=84mm]{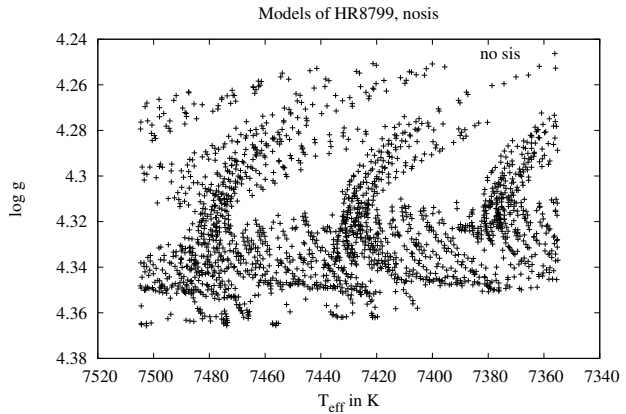}
 \caption{Position on the HR diagram of the models fulfilling the spectroscopic
   observations. The label ``nosis'' at the top of the figure means
   that no seismological constraints are used.}
\end{figure}

We have developed a grid of CESAM equilibrium models varying the mass
(in the range [1.25, 2.10] $\rm{M}_\odot$ with steps of $0.01
\rm{M}_\odot$), the metallicity (with values [M/H]=0.08, $-0.12$,
$-0.32$, and $-0.52$), the Mixing-Length parameter $\alpha_{\rm{MLT}}$
(values 0.5, 1, and 1.5), and the overshooting (values 0.1, 0.2, and
0.3).

Figure 4 shows in an HR diagram the models in our grid fulfilling the
observations. In Table 3, the range of free parameters of acceptable
models is shown. With this first set of observational constraints we
have reduced the number of acceptable models from around $600,000$ to
1975. This number is still large, but the first main result of this
study is that no models with metallicity larger than solar are in the
set of acceptable models. To study the reason for this, a new set of
models with solar metallicity has been computed. Most of these tracks
present models fulfilling all the constraints except the
luminosity. In Fig. 5 we present the luminosity as a function of the
effective temperature for tracks of different metallicities and
masses. All the solar metallicity models present luminosities larger
than the observed one for the effective temperatures adopted
(represented as a box in the figure).

We conclude, then, that an internal solar metallicity must be
discarded for this star. A luminosity 0.2 $L/L_{\odot}$ larger or a
lower effective temperature can make solar metallicity models
marginally acceptable. These would be young models located at the
ZAMS.

\begin{figure}
\includegraphics[width=84mm]{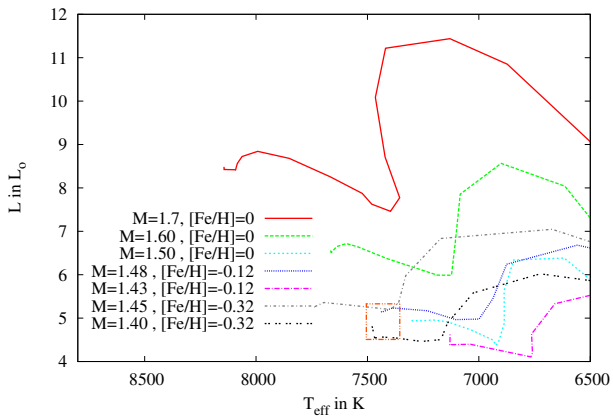}
 \caption{Effective temperature vs. stellar luminosity of some
   evolutionary tracks (from PMS to ZAMS) with different masses and
   metallicities, as described in the labels. The box represent the
   location of valid parameters for HR\,8799}
\end{figure}

The next goal of this study is to obtain an internal metallicity as
accurate as possible with the help of asteroseismology.

\subsection{Frequency Ratio Method}

The observation of three pulsational frequencies in the asymptotic
g-mode region makes the FRM suitable for the seismological study of
this star. First we must verify if the observed frequencies can be
part of the same rotation splitting (making the FRM not valid). This
verification is made by taking into account that the largest
difference among the observed frequencies is $3.8\,\mu \rm{Hz}$. On
the other hand, the first order approximation to the rotation
splitting provides a difference of the order of the rotation velocity,
weighted by a constant usually larger than 0.5 for $\ell=1$
  g-modes and 0.8 for $\ell=2$ g-modes \citep{chle}. In Figure 3, the
smallest rotation velocities, obtained for $i\approx 90^\circ$ , are
larger than 6.52 $\mu \rm{Hz}$. Therefore, the possibility of the two
more separate observed frequencies being part of the same splitting is
unlikely, since we would need the combination of a Ledoux constant
around 0.6 (only possible for some $\ell=1$ asymptotic g-modes)
  and, an inclination angle larger than $70^\circ$, a very unlikely
  value for this star taking into account the image obtained of this
  system, suggesting that the inclination angles are not close to
  equator-on rotation. This means that the first applicability
condition of the FRM is fulfilled.

The second verification of the applicability of the FRM is done on
the rotation velocity itself. \cite{frmrot} showed that, as the
analytical asymptotic expression is developed assuming no rotation,
the uncertainty in the application of the expressions grows with the
rotation velocity, until a limit where the method cannot be used (as
explained in section 3.3).

\begin{figure}
\includegraphics[width=84mm]{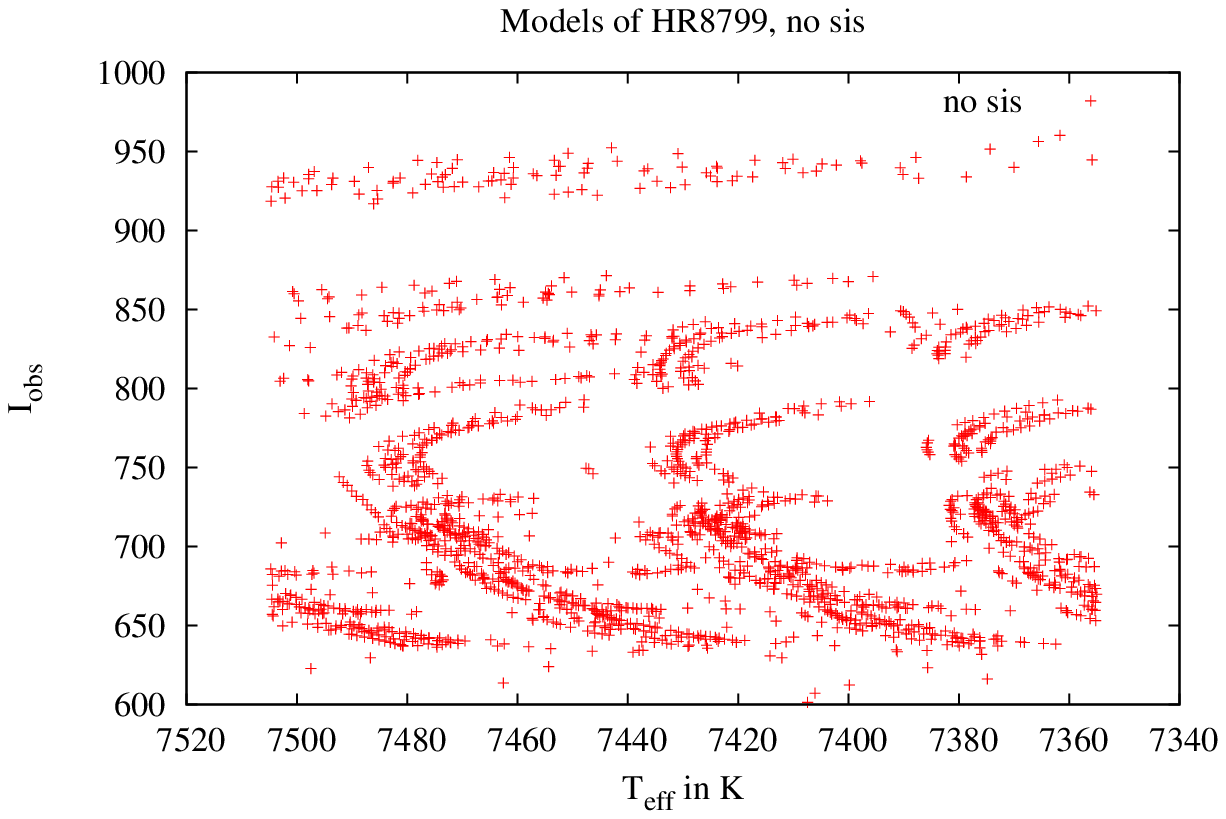}
\includegraphics[width=84mm]{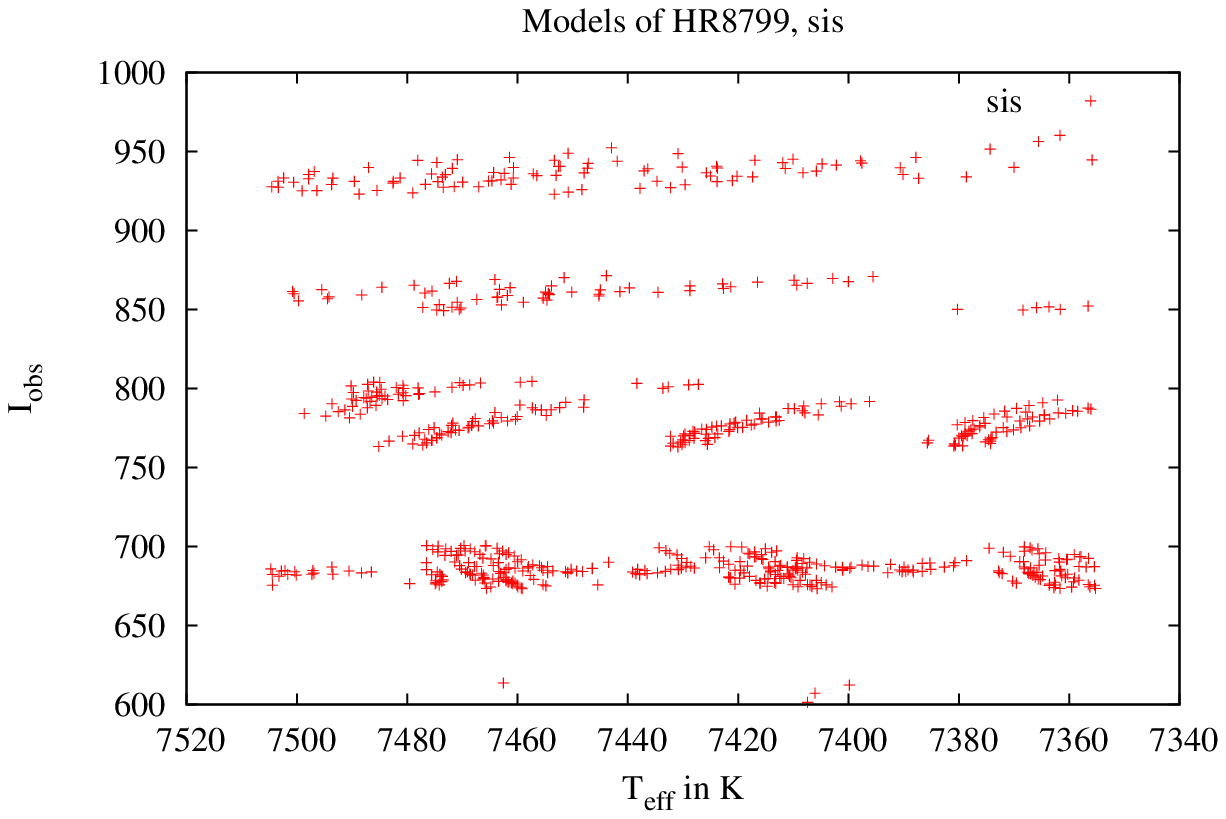}
 \caption{Effective temperature vs. Observed Brunt-V\"ais\"al\"a
   integral ($\rm{I}_{obs}$) obtained with the models fulfilling the
   spectroscopic constraints (upper panel, no seismological
   constraints used ``no sis'') and
   those restricted by the FRM (bottom panel, seismological
   constraints ``sis'').}
\end{figure}

One of the main assumptions of the FRM is the need that all the
frequencies used must have the same azimuthal order $m$. Currently
there is no method that can provide an univocal determination for the
value of $m$, the only one providing some information being optical
high-resolution spectroscopy. Therefore, we cannot verify this
assumption observationally, although there are some indications of the
fulfilling of this requirement: 1) The work of \cite{frmrot}, where
the case of modes with different $m$ is studied.  Their conclusion is
that the procedure is self-consistent. The FRM gives, in that case,
no-solutions, but not a wrong solution. This conclusion can be,
nevertheless, conditioned by the use of the perturbative
approximation. 2) The visibility of the modes with different $m$ in a
rotating star is described in \cite{visibility}. For the inclination
angles studied in this work, the visibility of the modes favours those
with $m=\pm 1$ in the case of the only mode identified for this star
($f_1$ having $\ell=2$, see next sections). Therefore, the only
possibility for the observed frequencies to have different m is to
have $m$=1 and $m=-$1. Taking into account the previous study of the
value of the rotational splitting, this means that the respective
$m$=0 are well separated and models can only marginally over-stabilize
both modes at the same time. In the unlikely case of quasi
  equator-on inclination angles, the visibility of the modes favours those
  with $m=\pm 2$, resulting in a larger separation of the
  corresponding $m=0$ modes and an even more unlikely
  over-stabilization of both modes at the same time. The only
possible scenario invalidating this assumption is, then, a $f_2$
having $\ell=1$ and m=0.

In this work we present two studies for two possible rotation
velocities. A first one for a moderate rotation velocity from the
point of view of the FRM, around 45 $\rm{km}\,\rm{s}^{-1}$, (this
  value has been chosen just for illustration purposes because it
  clearly shows how the procedure used here works). \cite{fabrycky}
  found that this rotation velocity cannot be excluded for HR\,8799,
  acording to the dynamical study of the system. In that case the
complete procedure gives very restrictive results regarding the number
of accepted models. At the end of the paper (Section 4.5) we show the
results obtained for the largest rotation velocity admitted by the FRM
(around 60 $\rm{km}\,\rm{s}^{-1}$), where this procedure is less
restrictive.

In the case of a rotation velocity of 45 $\rm{km}\,\rm{s}^{-1}$ we
have obtained a set of 10 chains, each one with a fixed set of mode
identifications and Brunt-V\"ais\"al\"a frequency integral. Fig. 6
shows the set of models fulfilling the additional constraint of
$\rm{I}_{obs}$ (bottom panel) as compared with those only fulfilling
the spectroscopic observations (upper panel). The number of rejected
models, for this moderate rotation velocity, is around $70\%$, but a
large number of models is still accepted (see Table 3), and only those
with overshooting 0.3 are rejected. This dependence between the
Brunt-V\"ais\"al\"a frequency and the overshooting parameter is well
known \citep{miglio} and is part of the valuable information provided
by the FRM.

\subsection{Instability analysis.}

The next step is to verify if the observed modes are predicted to be
over-stable for the remaining models, as explained in Section
3.4. Unfortunately, the TDC is not a complete description of the
pulsation - convection interaction, and the energy balance provided by
this study is not totally accurate. Therefore, we have adopted a
conservative criterion for the acceptance of a model: if the modes in
the observed frequency range present a growth rate positive or
negative but very close to zero, they are accepted. With this
criterion, some incorrect models will be considered as acceptable, but
we prefer this inconvenience instead of rejecting correct models.

This study is robust in selecting models, keeping only $10\%$ of those
filtered in the previous step (see Fig. 7 compared with Fig. 4). As
the energy balance is a phenomenon mainly thermodynamic, the accepted
models are located in a certain $T_{\rm{eff}}$ range, in the coolest
part of the HR diagram. This instability study has also rejected all
the models with metallicity $-0.52$, and $\alpha_{\rm{MLT}}$ 0.5 and
1.0 (as expected). The remaining valid masses are in the ranges
[1.32,1.33] and [1.44,1.45] $\rm{M}_{\odot}$.

\begin{figure}
\includegraphics[width=84mm]{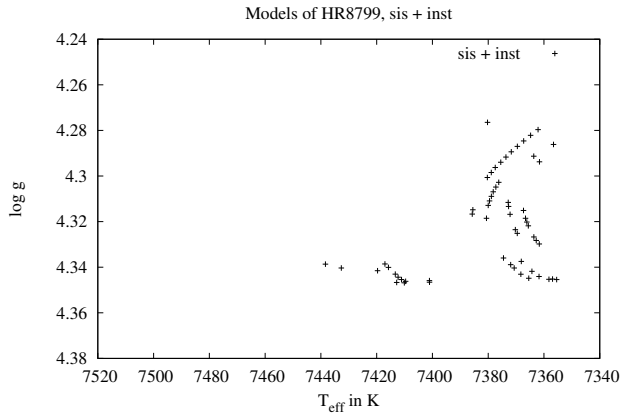}
 \caption{HR diagram position of the models fulfilling the spectroscopic
   observations + FRM + instability analysis.}
\end{figure}

\subsection{Mode identification with multicolour photometry.}

The last step of our study is to use the mode identification from the
multicolour photometry obtained for this star \citep{zerbi}. As
explained in Section 3.5, the non-adiabatic code, with the TDC
implemented, makes it possible to obtain accurate non-adiabatic
observables for these stars. With these observables we can carry out
the identification of the spherical degree $\ell$ of the observed
modes comparing the theoretical predictions for different modes with
observations. In Fig. 8, this comparison, in the phase differences
vs. amplitude ratio diagram, is done for a representative model. As
all the acceptable models at this stage are in a narrow range of
temperatures and masses, the rest of the models provide similar
results.

\begin{figure}
\includegraphics[width=84mm]{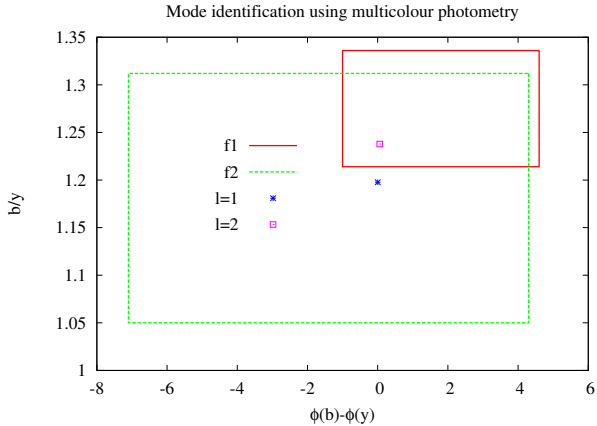}
 \caption{Phase differences vs. amplitude ratio between bands $b$
   and $y$ of the two observed frequencies with highest amplitude
   (boxes) as compared with theoretical predictions for different
   spherical degrees $\ell$ (points).}
\end{figure}

In Fig. 8, the observational errors are displayed as boxes, and
only the comparison between bands $b$ and $y$ is made. The
comparison of bands $v$ and $u$ with $y$ produce the same type of
behaviour. It can be easily checked that the mode $f_2$ cannot be
identified due to its large observational errors, but $f_1$ is always
identified as $\ell=2$. This is a strong constraint to the models,
since the FRM predicts fixed mode identifications for each particular
$\rm{I}_{\rm{obs}}$. Therefore, we have to select only those models
for which $f_1$ is predicted to have $\ell=2$. These models are shown
in Fig. 9.

\begin{figure}
\includegraphics[width=84mm]{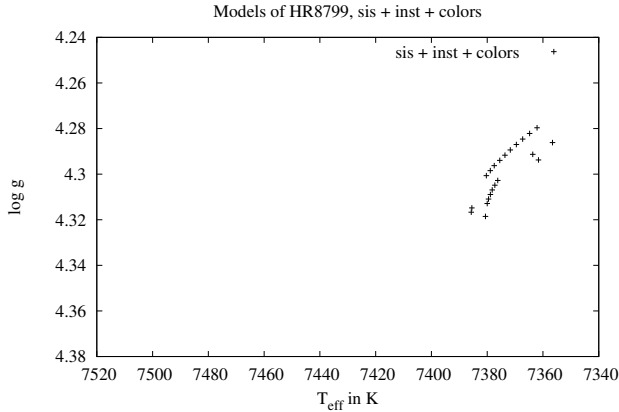}
 \caption{HR diagram position of the models fulfilling the spectroscopic
   observations + FRM + instability analysis + multicolour photometric
   mode identification.}
\end{figure}

The number of final models fulfilling all the observational
constraints is drastically decreased, remaining only 22 models of the
initial 1975 models selected not using asteroseismology. These 22
models have [M/H]=$-0.32\pm0.1$ (higher than that given by \cite{gray}
[M/H]=$-0.47\pm0.1$, but consistent with the error bars),
M=$1.32\pm0.01\,\rm{M}_{\odot}$, $\alpha_{MLT}=1.5\pm0.3$, and
overshooting $0.15\pm0.10$.

\subsection{The less favourable case}

We have already mentioned that the rotation velocity is one of the
main uncertainties for this procedure. All the previous studies have
been developed assuming a rotation velocity in the middle of the
applicability range of the method, that is, with a moderate rotation
velocity, or an inclination angle $i=50^\circ$-$60^\circ$. This may
not be the case. Therefore we have also repeated the same procedure
for the less favorable case: a rotation velocity around
$60\,\rm{km}\,\rm{s}^{-1}$, which implies an inclination angle around
$i\approx 36^\circ$. In this case, the most affected part of the
procedure is the FRM, increasing the errors and, as a consequence, the
number of acceptable models and $\rm{I}_{\rm{obs}}$. The final result
is not as detailed as for the moderate rotating case, and the final
set of models fulfilling all the constraints is displayed in
Fig. 10. The number of acceptable models is 66, with [M/H]=$-0.32$ and
$-0.12$, M=[1.32,1.33] and [1.44,1.45] $\rm{M}_{\odot}$,
$\alpha_{\rm{MLT}}=1.5$, and overshooting= 0.1, 0.2, 0.3. Analysing
the differences with the $v_{\rm{rot}}=45\,\rm{km}\,\rm{s}^{-1}$ case
we have noted that the main one is the inclusion of models with
overshooting 0.3, making it possible to fulfill the constraints
  the models with [M/H]=$-0.12$, and M=[1.44,1.45] $\rm{M}_{\odot}$.

\begin{figure}
\includegraphics[width=84mm]{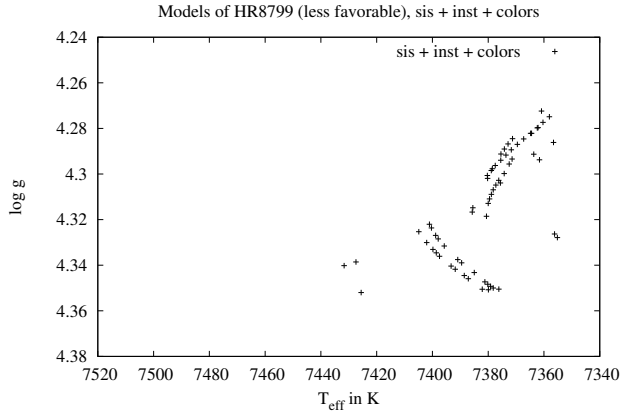}
 \caption{HR diagram position of the models fulfilling the spectroscopic
   observations + FRM + instability analysis + multicolour photometric
   mode identification, for the less favourable case in the FRM.}
\end{figure}

\section{Conclusions and perspectives}

In this work, the first comprehensive asteroseismologic study of the
planetary system host HR\,8799, a $\lambda$ Bootis star presenting
$\gamma$ Doradus pulsations has been carried out. This asteroseismic
work is specially important for the determination of the internal
abundances of this kind of stars, a previous step to understand the
physical mechanism responsible for the surface chemical peculiarities
of the $\lambda$ Bootis group. The asymptotic g modes observed in
HR\,8799 make it possible to deeply test its internal metallicity,
something hard to do with $\delta$ Scuti pulsators, for example.

For the asteroseismologic study, we have used part of the most updated
codes and physics for describing $\gamma$ Doradus stars: CESAM and
GraCo codes, TDC theory, FRM, and mode identification using
Str\"omgren multicolour photometry. We have constructed a dense grid
of equilibrium models, and carried out a first selection using the
available data ($T_{\rm{eff}}$, $\log\,\rm{g}$, and luminosity),
leaving the metallicity and mass as free parameters in order to test
the internal metallicity of the star.

The first selection shows that there are no models with solar
metallicity fulfilling the observations, since the stellar luminosity
derived from the observations is smaller than any of the possible
luminosities of models with solar metallicity. This contradicts the
main assumption of the theories explaining the $\lambda$ Bootis
nature, i.e. that these stars have solar metallicity, whereas the
observed abundances are due to surface phenomena.

The asteroseismic study to fix the metallicity of the star is done
using 1) the FRM, to estimate possible mode identifications and
Brunt-V\"ais\"al\"a integrals, 2) energy balance to determine the
stability of the modes, selecting models in a certain temperature
range and $\alpha_{\rm{MLT}}$ and, 3) mode identification with
multicolour photometric observables, selecting only models for which
FRM predicts the mode with the highest amplitude to have
$\ell=2$. This is the first time that such a study has been done for a
planet-hosting star, but the general procedure was proposed in
\cite{completo}. All these steps provide a small group of acceptable
models depending on the rotation velocity. For a moderate rotation
velocity (implying $i$ around $50^\circ$) the mass and the metallicity
are very precise determined ([M/H]=$-0.32\pm0.1$ and M=$1.32\pm
0.01\rm{M}_{\odot}$). For a rotation velocity at the limit of the
validity of the FRM (around $i\approx 36^\circ$), the method is less
selective and provides some possible values of the metallicity and
mass ([M/H]=$-0.32$ and $-0.12\pm0.1$, and M=[1.32,1.33] and
[1.44,1.45] $\rm{M}_{\odot}$ respectively). We point out that
inclination angles lower than $18^\circ$ cannot be possible to
preserve the $\lambda$ Bootis nature of the star. Nevertheless, this
limit must be also verified with more accurate observations.

These results have some consequences. The first one is related to the
most accepted theory of the $\lambda$ Bootis nature, namely the
accretion/diffusion scenario given by \cite{VennLambert90}. The result
of this work, discarding solar metallicity as the internal metallicity
of the star, together with the fact that some $\lambda$ Bootis stars
have debris disks not connected with the star \citep{chen}, makes this
explanation unlikely, but not negligible \citep{su}. In any case, the
equation developed by \cite{VennLambert90} to describe the individual
abundances as a sum of internal abundance plus that of accreted
material, must be corrected to include a possible non-solar internal
abundance as following

\begin{equation}
\epsilon(m)=(1-f)\epsilon_*(m)+f\epsilon_{\rm{ISM}}(m)
\end{equation}

\noindent where $f$ is a mixture factor, and $\epsilon$, $\epsilon_*$
and $\epsilon_{\rm{ISM}}$ are the abundances observed, internal and coming
from the interstellar medium of each chemical element,
respectively. Therefore, the study of internal chemical mixing
processes seems to be the key to explain the $\lambda$ Bootis nature,
at least for HR\,8799, as the solar abundances for C, N, O and S
observed on its surface have still to be explained.

On the other hand, we have seen that the asteroseismological study
needs an accurate determination of the rotation velocity, not just the
projected rotation velocity, of the star, for a very accurate
determination of the stellar characteristics. Therefore, a consequence
of this study is the need for a precise determination of the
inclination angle $i$, of the multicolour photometric amplitudes and
phases of $f_2$, and some information of $m$ values through
time-series if high resolution spectroscopy. These determinations
would help to carry out a definitive selection of the models.

\section*{Acknowledgments}

The authors want to thank an anonymous referee for his/her useful
suggestions and comments. AM acknowledges A. Moro-Mart\'in her help
and, the financial support from a Juan de la Cierva contract of the
Spanish Ministry of Science and Innovation. PJA acknowledges financial
support from a "Ramon y Cajal" contract of the Spanish Ministry of
Education and Science. This research has been funded by Spanish grants
ESP2007-65475-57-C02-02, CSD2006-00070 and ESP2007-65480-C02-01.

\end{document}